\documentclass[prc,letterpaper,twoside,superscriptaddress,showpacs]{revtex4}

\usepackage[bookmarks,dvips,pdfhighlight=/O,pdfstartview=FitH]{hyperref}
\usepackage{graphicx,amsmath,latexsym,amssymb,mathrsfs,amsopn,bm,mathbbol,qmech-revtex}
\usepackage[dvips]{color}
\begin{document}

\title{Delta(1232) and Nucleon Spectral Functions in Hot Hadronic
    Matter}

\author{H. van Hees}
\author{R. Rapp}
\affiliation{Cyclotron Institute and Physics Department, Texas A\&M 
University, College Station, Texas 77843-3366}
\date{\today}

\begin{abstract}
  Modifications of $\Delta(1232)$ and nucleon spectral functions at finite
  temperature and baryon density are evaluated in terms of resonant
  scattering off thermal pions, as well as a renormalization of the vacuum
  $\Delta$-width including vertex corrections. The interactions are based
  on effective Lagrangians of pions and baryon resonances, with underlying
  parameters (coupling constants and form factors) determined by the
  elastic $\pi N$ scattering phase shift in the isobar channel, and by
  empirical decay branchings of excited resonances.  Pion modifications are
  included via interactions in a pion gas, as well as standard $\Delta$-
  and nucleon-hole excitations in nuclear matter. In hot hadronic matter,
  under conditions resembling thermal freezeout at the Relativistic
  Heavy-Ion Collider (RHIC), the $\Delta$ exhibits a significant broadening
  of $\sim$65~MeV together with a slight upward peak shift of 5-10~MeV, in
  qualitative agreement with preliminary data from the STAR collaboration.
\end{abstract}

\maketitle

%%%%%%%%%%%%%%%%%%%%%%%%%%%%
\section{Introduction}
%%%%%%%%%%%%%%%%%%%%%%%%%%%%%%%
In strong interactions at low and intermediate momentum transfer the
relevant degrees of freedom are associated with colorless hadrons. Whereas
their interactions are constrained by the approximate chiral symmetry (CS)
in the two-flavor sector of Quantum Chromodynamics (QCD), their mass
spectrum is largely governed by the spontaneous breaking of CS (SBCS) in
the QCD vacuum.  At temperatures $T\simeq$~150-200~MeV, CS is expected to
be restored~\cite{kar01}, implying substantial modifications of the
hadronic spectrum close to the critical temperature. Pertinent medium
modifications of hadrons are hoped to be identified by creating hot and
dense nuclear matter in high-energetic collisions of heavy nuclei.

In recent years, intense theoretical efforts have been devoted to
understand in-medium properties of especially vector mesons, which are
rather directly accessible in heavy-ion experiments through decays into
dileptons, emitted throughout the hot and dense phases of the collision
with negligible final-state interactions (for a review, see, e.g.,
Ref.~\cite{rw99}). Baryonic effects have been found to be of particular
importance in accounting for the large excess production of lepton pairs,
observed in Pb-Au collisions~\cite{ceres98,ceres03}, at invariant masses
below the vacuum $\rho$-mass.  Consequently, towards a more complete
picture, in-medium effects of the baryon spectrum itself deserve further
investigation.

In the present article we calculate modifications of the nucleon ($N$) and
$\Delta(1232)$ particles in a hot and dense medium within an effective
quantum-field theoretical hadronic model coupled with standard many-body
techniques. Besides their (indirect) role in dilepton production, $\Delta$
properties in heavy-ion collisions are accessible via resonance
spectroscopy, i.e., mass and width changes in the line shapes of $\pi N$
invariant-mass spectra~\cite{hjo97,pelte97,zhang04,fach04a}. 
In-medium $\Delta$ spectral function to date were mostly evaluated 
for cold nuclear matter~\cite{hori80,os87,ew88,mig90,xia94,lutz03},
with few exceptions~\cite{ko89,korpa95}. Here we go beyond the latter works
by performing a more detailed treatment of its pion-nucleon cloud at finite
temperature including vertex corrections, as well as resonance excitations
induced by thermal pions. Also, thermal self-energies of the nucleon beyond
the $\Delta$ contribution are, to our knowledge, assessed for the first 
time~\cite{LS90,DLR94,KZ96,EI97}.  In a more general context, the change of the
baryon spectrum towards chiral restoration is an important component in the
understanding of QCD phase transitions in the $\mu_B$-$T$ plane ($\mu_B$:
baryon chemical potential).

Our article is organized as follows: In Sec.~\ref{sec_vac} we introduce the
hadronic interaction Lagrangians that are subsequently employed and discuss
the determination of the free parameters, using vacuum scattering and decay
data. In Sec.~\ref{sec_self} we compute in-medium
self-energies of nucleon and $\Delta$. The latter are used in
Sec.~\ref{sec_spec} to evaluate $N$ and $\Delta$ spectral properties in
matter. For the $\Delta$, we first qualitatively confront the effects of
cold nuclear matter with information from nuclear photoabsorption, and then
discuss both $N$ and $\Delta$ spectral functions in hot and dense matter
under conditions expected for heavy-ion experiments at various collision
energies.  In Sec.~\ref{sec_concl} we summarize and give an outlook.

%%%%%%%%%%%%%%%%%%%%%%%%%%%%%%%%%%%%%%%%%%%%%%%%%%%%%%%%%%
\section{Hadronic Interaction Lagrangians in Vacuum}
\label{sec_vac}
%%%%%%%%%%%%%%%%%%%%%%%%%%%%%%%%%%%%%%%%%%%%%%%%%%%%%%%%%%%%%
The central quantities of interest in our present analysis are the
in-medium propagators of the two lowest-lying non-strange baryon states,
nucleon and $\Delta$.  Throughout this article we employ a
quasi-relativistic description of the baryon fields, i.e., use relativistic
dispersion relations, $E_B^2(\vec{p})=m_B^2+\vec{p}^2$ ($B$=$N$, $\Delta$,
$N^*$, $\Delta^*$), but neglect anti-particle contributions and
additionally restrict Rarita-Schwinger spinors to their non-relativistic
spin-3/2 components. Pions are treated fully relativistic, with
$\omega_\pi^2(\vec k)= m_\pi^2 +\vec{k}^2$. Within this approximation, the
free retarded propagators for baryons and pions read
\begin{equation}
\label{1}
G_{B}^{(0)}(p) = \frac{1}{p_0-E_{B}(\vec{p})-\Sigma_B^{(0)}(p)} \ , \qquad
G_{\pi}^{(0)}(k) = \frac{1}{k_0^2 - \omega_{\pi}^2(\vec{k}) + 
\ii \eta \sign(k_0)},
\end{equation} 
where $\Sigma_B ^{(0)}(p)$ denotes the vacuum self-energy encoding the
partial decay branchings of each resonance. Except for the $\Delta$ (for
which a renormalization of its bare mass is accounted for via the $\pi N$
loop), we use constant physical pole masses implying that only the
imaginary part of $\Sigma_B^{(0)}$ is nonzero (for the nucleon,
$\Sigma_B^{(0)}$=0).

A key ingredient in our description are the interaction vertices involving
a pion and two baryons. The following notation is adopted below:
two-component spin-$1/2$ fields are denoted by $\psi$ and four-component
spin-$3/2$ fields by $\Psi$; the isospin of a resonance is indicated by an
appropriate subscript (``1'' and ``3'' for $I=1/2$ and $3/2$,
respectively). The dominant coupling type is usually given by the lowest
angular momentum in the pion decay (except for $\pi NN$, where $s$-wave
interactions are neglected), but higher waves are included whenever
empirically significant. Pions are represented by a real isospin triplet
$\vec{\phi}$ transforming according to the fundamental $\text{SO}(3)$
representation of the isospin group.  This leads us to the following
interactions~\cite{os87,cub90,rubw98,ubw99,novr01} involving nucleons,
\begin{align}
\Lag_{s,N}^{(11)} &= -f \psi_{1}^{\dagger} \vec{\tau} \vec{\phi} \psi_{N} +
\text{h.c.} \label{2a} \\
\Lag_{p,N}^{(11)} &= -\frac{f}{m_{\pi}} \psi_{1}^{\dagger} 
[(\vec{\sigma} \vec\nabla)
(\vec{\tau} \vec{\phi})] \psi_{N} + \text{h.c.} \label{2b} \\
\Lag_{s,N}^{(31)} &= -f \psi_{3}^{\dagger} \vec{T} \vec{\phi} \psi_N +
\text{h.c.}, \label{2c} 
\end{align}
and $\Delta$(1232) resonances, 
\begin{align}
\Lag_{s,\Delta}^{(13)} &= \ii g \Psi_{1}^{\dagger} \vec{T}^{\dagger} \vec{\phi} \Psi_{\Delta} +
\text{h.c.} \label{2d} \\
\Lag_{s,\Delta}^{(33)} &= \ii g \Psi_{3}^{\dagger} 2 \vec{\Theta} \vec{\phi}
\Psi_{\Delta}, \label{2d-2}\\
\Lag_{p,\Delta}^{(13)} &= -\frac{f}{m_{\pi}} \psi_{1} [(\vec{S}^{\dagger}
\vec\nabla)(\vec{T}^{\dagger} \vec{\phi})] \Psi_{\Delta} + \text{h.c.}\label{2e} \\
\Lag_{p,\Delta}^{(33)} &= -\frac{f}{m_{\pi}} \Psi_{3} [(2 \vec{\Sigma}
\vec{\nabla})(2\vec{\Theta} \vec{\phi})] \Psi_{\Delta} +\text{h.c.}\label{2f} \\
\Lag_{d,\Delta}^{(13)} &= -\ii \frac{f}{m_{\pi}^2} \psi_{1} [(\vec{S}
\vec\nabla)(\vec{S}^{\dagger} \vec\nabla) (\vec{T}^{\dagger} \vec{\phi})]
\Psi_{\Delta}+\text{h.c.} \label{2g} \\
\Lag_{d,\Delta}^{(33)} &= -\ii \frac{f}{m_{\pi}^2} \psi_{3}^{\dagger} [(\vec{\sigma} \vec\nabla)
(\vec{S}^{\dagger} \vec\nabla) (2 \vec{\Theta} \vec{\phi})] \Psi_{\Delta} +
\text{h.c.} \label{2g-2}
\end{align}
In the above equations, the superscripts on the left-hand-side refer to the
isospin content of the two baryon fields, whereas the subscripts $l,N$
($l,\Delta$) indicate the angular momentum $l$=$s$, $p$ and $d$ in the
$\pi N$ ($\pi\Delta$) system.  Furthermore, $\vec{\sigma}$ ($\vec{\tau}$)
are Pauli matrices in (iso-) spin space, and $\vec{S}$ ($\vec{T}$) the
pertinent 1/2$\to$3/2 transition matrices (consisting of appropriate
Clebsch-Gordan coefficients which project the 1-1/2 couplings onto pure 3/2
states). Finally, $\vec{\Sigma}$ ($\vec{\Theta}$) denote the (iso-)
spin-3/2 (4$\times$4-) matrices.  The finite size of the vertices is
accounted for by phenomenological hadronic form factors, which are of
monopole type for $s$- and $p$-wave couplings (\ref{2a}-\ref{2f}),
$F_{\text{mon}}(|\vec{k}|)=\Lambda^{2}/(\vec{k}^2+\Lambda^2)$, and of
dipole type for the $d$-wave couplings (\ref{2g}) and (\ref{2g-2}),
$F_{\text{dip}}(|\vec{k}|)=4 \Lambda^4/(2 \Lambda^2+\vec{k}^2)^2$.

Except for $\pi NN$ and $\pi N\Delta$ vertices, all $\pi B_1B_2$ coupling
constants are adjusted to the (average) empirical values of the
corresponding partial $B_2\to B_1 \pi$ decays (using pole masses and
widths) according to the Particle Data Group~\cite{pdb02}, together with a
uniform form-factor cutoff $\Lambda_{\pi \Delta B}$=500~MeV, cf.
Tab.~\ref{tab_1}.
\begin{table}
\begin{tabular}{|l|l|l|}
\hline
Vertex & Eq.  & $f$ ($g$)\\
\hline
$\pi N N$ & (\ref{2b}) & 1.0 \\
$\pi N \Delta$ & (\ref{2e}) & 3.2  \\
$\pi N N(1440)$ & (\ref{2b}) & 0.779  \\
$\pi N N(1535)$ & (\ref{2a}) & 1.316  \\
$\pi N \Delta(1600)$ & (\ref{2e}) & 1.170  \\
$\pi N \Delta(1620)$ & (\ref{2d}) & 0.828 \\
\hline 
$\pi \Delta N(1440)$ & (\ref{2e}) & 2.185  \\
$\pi \Delta \Delta(1600)$ & (\ref{2f}) & 0.211  \\
$\pi \Delta N(1520)$ & (\ref{2d}) & 0.760  \\
                     & (\ref{2g}) & (-1.126) \\
$\pi \Delta N(1700)$ & (\ref{2g}) & 0.351  \\
$\pi \Delta \Delta(1620)$ & (\ref{2g-2}) & 0.111 \\
$\pi \Delta \Delta(1700)$ &(\ref{2d-2}) & 0.655 \\
\hline
\end{tabular}
\caption{$\pi NB$ and $\pi\Delta B$ vertices and coupling constants
used in the present analysis.} 
\label{tab_1}
\end{table}
Since the main focus in the present article is on spectral properties of
the $\Delta$ resonance, we calculate its vacuum self-energy including a
finite real part. With the interaction vertex, Eq.~(\ref{2f}), and using
free pion and nucleon propagators, its imaginary part takes the form
\begin{equation}
\im \Sigma_{\Delta}^{(N\pi)}(M) = -\frac{f_{\pi N \Delta}}{12 m_{\pi}^2 \pi}
\frac{m_N k_{\text{cm}}^3}{M} F^2(k_{\text{cm}}) \Theta(M-m_N-m_{\pi}),
\quad  k_{\text{cm}}^2=\frac{(M^2-m_N^2-m_{\pi}^2)^2-4m_N^2m_\pi^2}{4M^2} \ .
\label{gd_vac}
\end{equation}
Here, we introduced an extra factor $m_N/E_N(k_{\text{cm}})$ to ensure a
Lorentz-invariant decay width. The real part of the self-energy is then
determined by its spectral representation. Assuming the $\pi N$ interaction
in the 33-channel to be dominated by the $\Delta$ resonance allows to
relate the self-energy to the $\pi N$ elastic scattering phase shifts via
\begin{equation}
\tan[\delta_{33}(M)] = \frac{\im G_{\Delta}(M)}{\re G_{\Delta}(M)}.  
\end{equation} 
As is well known, within this approximation a satisfactory fit to the 
$\delta_{33}$ phase shift~\cite{mon80,korpa95,wfn98} requires a low 
cutoff, $\Lambda_{\pi N \Delta}$=290~MeV, together with a 
large coupling constant, $f_{\pi N\Delta}$=3.2, and a bare 
mass of $m_{\Delta}^{(0)}$=1302~MeV~\cite{mon80,korpa95,wfn98}. 
The same cutoff will be employed for the $\pi NN$ vertex with the 
standard value for the coupling constant, $f_{\pi NN}$=1.

Baryon-baryon interactions via $t$-channel meson exchange are negelcted in
the present analysis. In high-temperature matter, which is our main
interest here, we expect resonant meson-baryon scattering to be more
important, as it was the case for earlier calculations of the pion and
$\rho$-meson spectral functions in hot and dense matter~\cite{rw99}.

%%%%%%%%%%%%%%%%%%%%%%%%%%%%%%%%%%%%%%%%%%%%%%%%%%%%%%%%%%%%%
\section{In-Medium Self-energies}
\label{sec_self}
%%%%%%%%%%%%%%%%%%%%%%%%%%%%%%%%%%%%%%%%%%%%%%%%%%%%%%%%%%%%%%%
With the above interaction vertices we proceed to evaluate
$N$ and $\Delta$ self-energies in hot and dense hadronic matter.

The in-medium $\Delta$ self-energy has two components. The first, and more
involved one is the $\pi N$ decay extended to finite temperatures and
baryon densities. We employ the imaginary-time (Matsubara) formalism to
calculate the corresponding $N \pi$-loop diagram according to
\begin{equation}
\begin{split}
\label{sgdnpi}
  \Sigma_{\Delta}^{(N\pi)}(p) &= -\ii \frac{f_{\pi N \Delta}^2}{3 m_{\pi}^2}
  \int \frac{d^3 \vec{l}}{(2\pi)^3} \frac{m_N}{E_N(\vec{l})} T \sum_{z_\nu}  F^2 (|\vec{k}|)
  \vec{k}^2 G_{\pi}(\ii \omega_{\kappa}-\ii
  z_{\nu},\vec{p}-\vec{k}) G_N(\ii z_{\nu},\vec{l})  \\
  &= \frac{f_{\pi N \Delta}^2}{3 m_{\pi}^2} \int \frac{\d^4 l}{(2 \pi)^4}
  \frac{m_N}{E_N(\vec{l})} \vec{k}^2 F_{\pi}^2(\vec{k}^2) \lbrace
  [\Theta(k_0)+\sigma(k_0) f^{\pi}(|k_0|)]
  {A}_{\pi}(k) G_N(l) 
-f^N(l_0) A_N(l) G_{\pi}(k) \rbrace \ ,
\end{split}
\end{equation}
where $k=p-l$ is the pion four-momentum, and $A_{N}=-2\im G_{N}$ and
${A}_{\pi}=-2 \im {G}_{\pi}$ denote the in-medium nucleon and pion spectral
function, respectively (the extra factor $m_N/E_N(\vec{l})$ ensures
consistency with Eq.~(\ref{gd_vac})).  The thermal distribution functions
are given by $f^{N}(l_0)=f^{\text{fermi}}(l_0-\mu_N,T)$ and
$f^{\pi}(|k_0|)=f^{\text{bose}}(|k_0|,T) \exp(\mu_{\pi}/T)$, with
$f^{\text{fermi}}$ and $f^{\text{bose}}$ being the standard Fermi and Bose
functions, respectively.  To avoid Bose poles in the presence of finite
widths for in-medium pion spectral functions, pion-chemical potentials,
$\mu_{\pi}$$>$0, are treated in Boltzmann approximation.

\begin{figure}[t]
\begin{center}
\parbox{0.3\linewidth}{\includegraphics[width=\linewidth]{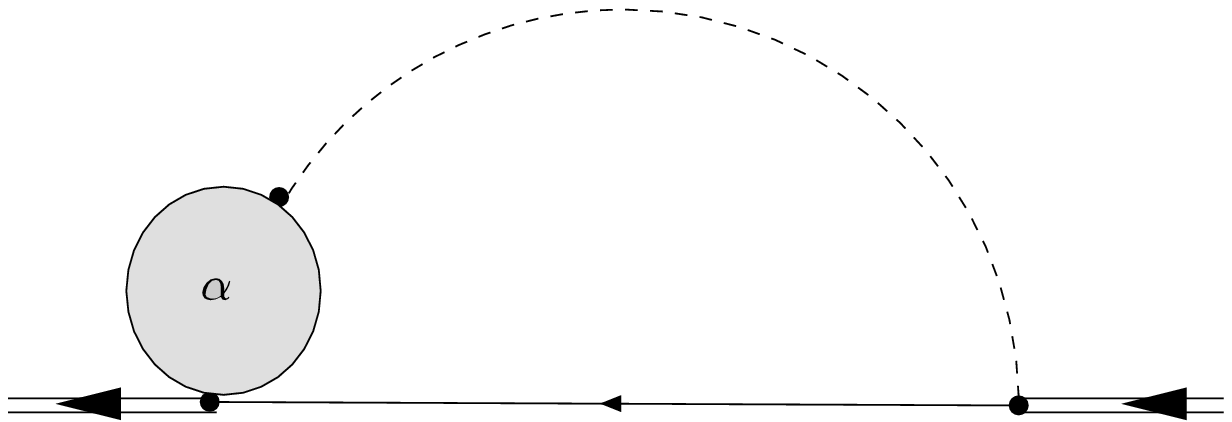}}\hspace*{2mm}
\raisebox{-0.41mm}{\parbox{0.3\linewidth}{\includegraphics[width=\linewidth]{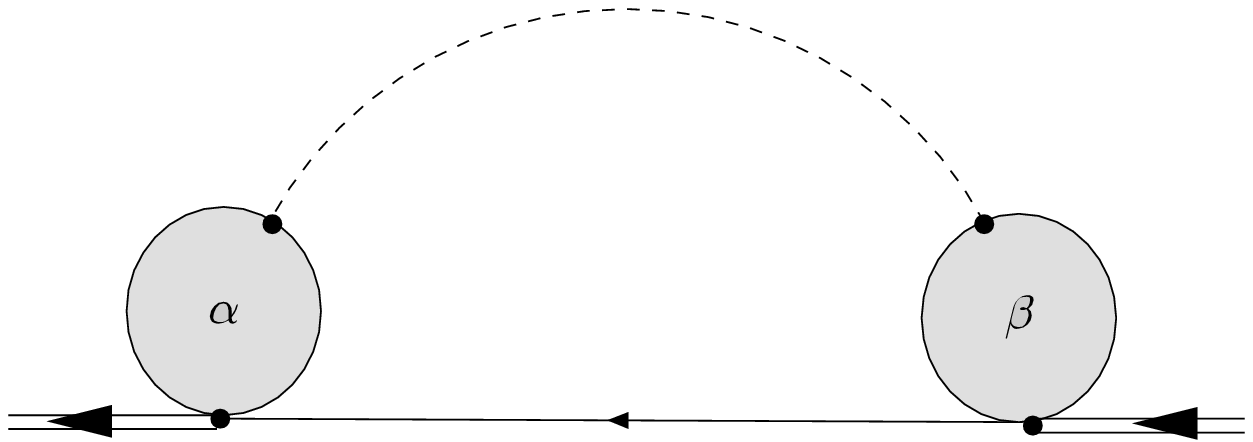}}}\hspace*{2mm}
\raisebox{-7.0mm}{\parbox{0.25\linewidth}{\includegraphics[width=\linewidth]{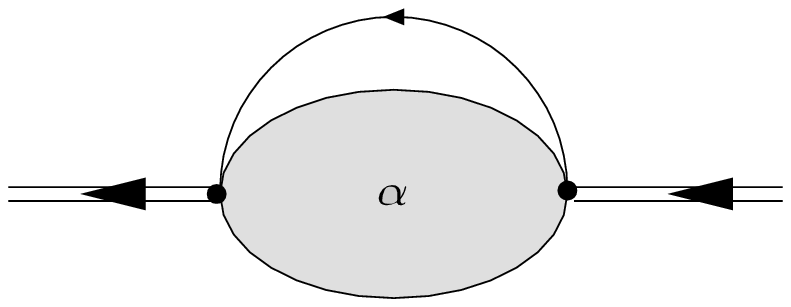}}}
\end{center}
\caption{Diagrammatic representation of the $\pi N\Delta$-vertex corrections
  (dashed lines: pion propagator, solid lines: nucleon propagator, double
  solid lines: $\Delta(1232)$ propagator); a blob with label $\alpha$
  corresponds to the Lindhard function $\Pi_{\alpha}$ ($\alpha \in
  \{1,2\}$) attached to the baryon lines with the pertinent Migdal
  parameters, i.e., $g_{12}'$ for $\alpha$$=$1 and $g_{22}'$ for
  $\alpha$$=$2.}
\label{fig_vcorr}
\end{figure}

The second equality in Eq.~(\ref{sgdnpi}) follows by using the Lehmann
representation for the propagators together with an analytical continuation
in the energies. Positive energies $k_0>0$ then correspond to outgoing
pions encoding the in-medium $\Delta\to\pi N$ decay, whereas contributions
with $k_0<0$ correspond to scattering with pions from the heat bath.  The
nucleon spectral function will be discussed in more detail below. The pion
spectral function is based on a pion self-energy
$\Sigma_\pi(k;T,\mu_\pi,\mu_N)$ that includes contributions from both
finite temperatures and baryon densities.  The temperature modifications
are modeled by a four-point $\pi\pi$ interaction to second order in the
coupling constant, corresponding to so-called sunset
diagrams~\cite{vHK2001-Ren-II}. The value of the coupling constant is
adjusted to approximately reproduce pion gas self-energies obtained from
more realistic $\pi\pi$ interactions in $s$-, $p$- and
$d$-wave~\cite{rw95}.  The baryonic effects are calculated in terms of
standard Lindhard functions for $p$-wave nucleon- and $\Delta$-hole
excitations at finite temperature including short-range correlations
parametrized by Migdal parameters~\cite{mig78} (due to the soft form
factors we use rather small default values $g_{11}'$=0.8 and
$g_{12}'$=$g_{22}'$=0.33~\cite{rubw98}). As is well-known from similar
evaluations of $\rho$-meson self-energies~\cite{cs93,hfn93} (see also
Ref.~\cite{ko89} for an early application for the $\Delta$), the $p$-wave
softening of the pion dispersion relation can induce an artificial
threshold enhancement due to a near vanishing of the pion group velocity.
This is remedied through the inclusion of vertex corrections, which, for
the vector-meson case, are, in fact, required to maintain a conserved
vector current.  Here, we adopt a similar procedure involving four-point
baryon vertices which are resummed in complete analogy to (and with the
same coupling constants as) the Migdal corrections in the pion propagator.
The vertex corrections amount to replacing the pion propagators in
Eq.~(\ref{sgdnpi}) by
\begin{equation}
\tilde{G}_{\pi}(k) =  G_{\pi}(k) \{ 1 + g_{12}' \Pi_1(k)
+ g_{22}' \Pi_2 + [g_{12}' \Pi_1(k)  +
g_{22}' \Pi_2]^2 \} 
+ \frac{{g'}_{12}^2 \Pi_1(k) + {g'}_{22}^2
\Pi_2(k)}{\vec{k}^2} \  , 
\label{vcorr}
\end{equation}
where $\Pi_1$ and $\Pi_2$ denote resummed finite-temperature Lindhard
functions corresponding to Feynman diagrams with outgoing $NN^{-1}$-, $N
\Delta^{-1}$-loops, respectively, cf. Fig.\ref{fig_vcorr}.
The vertex corrections are, in fact, closely related to the ``induced
interaction" as implemented in the nuclear $\Delta$ width in
Ref.~\cite{os87}. Here, we restrict ourselves to the longitudinal (pion-)
component; the transverse part including exchange of in-medium $\rho$
mesons will be addressed in future work.

The second component of the in-medium $\Delta$ self-energy arises from
interactions with thermal pions, approximated by $s$-channel resonance
formation.  Evaluating the finite temperature part of pertinent $\pi B$
loop diagrams within the Matsubara formalism leads to
\begin{equation}
\label{sgDpiB}
\Sigma_{\Delta}^{(\pi B)}(p)
        =-\int \limits_{l_0 \geq p_0}  \frac{\d^4 l}{(2 \pi)^4}
        v_{\pi\Delta B}^2(\vec{k}) 
\left[ f^{\pi}(|k_0|) 
A_{\pi}(k) \; G_B(l) + f^B(l_0) A_{B}(l) \; G_{\pi}(k) \right] \ .
\end{equation}
As before, $k$=$p-l$ is the pion-four momentum, and $v_{\pi NB}$ are the
interaction vertices (including hadronic form factors) following from the
Lagrangians (\ref{2d})-(\ref{2g-2}), summed (averaged) over spins and
isospins of the baryon resonance $B$ and the pion ($\Delta$).  In the
baryon-resonance Green's function in-medium effects on the decay width are
accounted for through thermal occupation factors, whereas modifications of
the real part (due to medium effects on the various decay products), as
well as possible vertex corrections, are neglected (recall that these two
effects tend to compensate each other, cf.  the discussion before
Eq.~(\ref{vcorr}) above).  Also note that we suppressed the contributions
of the $\pi B$ loop that survive in the vacuum (technically, this is
achieved by introducing the factor $\Theta(k_0)$ into the integrand). This
is to ensure consistency with our fit to the free $\pi N$ scattering phase
shifts where such diagrams have not been included (those would not only
affect the imaginary part of $\Sigma_\Delta^{(0)}$ above threshold, but
also require additional mass and wave-function renormalizations through
its real part).

The medium effects on the nucleon are attributed to $s$-channel
interactions with thermal pions. Starting from the equivalent expression
for the $\Delta$, Eq.(\ref{sgDpiB}), we apply the same level of
approximation as for the pions (encoded in the Lindhard functions), i.e.,
neglecting off-shell energy-dependencies in the thermal distribution
functions and resonance widths. For the $p$-wave scattering $\pi N$$\to$$N$
and $\pi N$$\to$$\Delta$, e.g., this leads to
\begin{equation}
\begin{split}
  \Sigma_N^{(T)}(p) = -\int \d^3 \vec{l} \ \frac{\vec{k}^2
    F_{\pi}^2}{\omega_{\pi}(\vec{k})} \Bigg \{ 
  \frac{3f_{\pi NN}^2}{16 \pi^3m_{\pi}^2}
  \frac{f^\pi[\omega_{\pi}(\vec{k})]
    +f^N[E_N(\vec{l})]} {E_N(\vec{l})-\omega_{\pi}(\vec{k})
    - p_0-\ii \epsilon}
   + \frac{f_{\pi N \Delta}^2}{12 \pi^3 m_{\pi}^2}
  \frac{f^\pi[\omega_{\pi}(\vec{k})]+f^\Delta[E_{\Delta}
    (\vec{l})]}{E_{\Delta}(\vec{l}) -\omega_{\pi}(\vec{k}) -
    p_0-\ii \Gamma_{\Delta}/2} \Bigg \} \ .
\label{sgNpi}
\end{split}
\end{equation}
and likewise for the other resonances listed in Table~\ref{tab_1}.

Let us finally comment on the chemical potentials entering the thermal
distribution functions. In high-energy heavy-ion collisions, hadron yields
can be rather accurately described by a ``chemical
freezeout''~\cite{pbm03}, characterized by a common temperature and baryon
chemical potential (depending on collision energy), with meson-chemical
potentials equal to zero. In subsequent hadronic cooling, finite chemical
potentials for stable mesons are required to maintain the observed hadron
ratios~\cite{rap02}.  Relative equilibrium for strong processes, e.g.,
$\pi\pi\leftrightarrow \rho$ or $\pi N\leftrightarrow \Delta$, then implies
relations of the type $2\mu_\pi$=$\mu_\rho$, $\mu_N+\mu_\pi$=$\mu_\Delta$,
which will be incorporated below.

%%%%%%%%%%%%%%%%%%%%%%%%%%%%%%%%%%%%%%%%%%%%%%%%%%%%%%%%%%%%
\section{\texorpdfstring{Nucleon and $\Delta$ Spectral Functions in 
   Medium}{Nucleon and Delta Spectral Functions in the Medium}}
\label{sec_spec}
%%%%%%%%%%%%%%%%%%%%%%%%%%%%%%%%%%%%%%%%%%%%%%%%%%%%%%%%%%%%%
%%%%%%%%%%%%%%%%%%%%%%%%%%%%%%%%%%%%%%%%%%%%%%%%%%%%%%%%%%%%
\subsection{Cold Nuclear Matter}
\label{ssec_cold}
%%%%%%%%%%%%%%%%%%%%%%%%%%%%%%%%%%%%%%%%%%%%%%%%%%%%%%%%%%%%
A valuable model test of the nuclear medium effects on the $\Delta$ can 
be performed by comparing its spectral properties to photoabsorption 
data on nuclei (see, e.g., Ref.~\cite{korpa04} for a recent example). 
This, in particular, allows to better constrain the in-medium 
modifications of the pion (depending on the $\pi NN$ form factor and 
Migdal parameters) and the impact of the $\pi N\Delta$ vertex 
corrections.
\begin{figure}[!t]
\begin{minipage}[b][8.5cm][t]{0.45\textwidth}
\includegraphics[width=\textwidth]{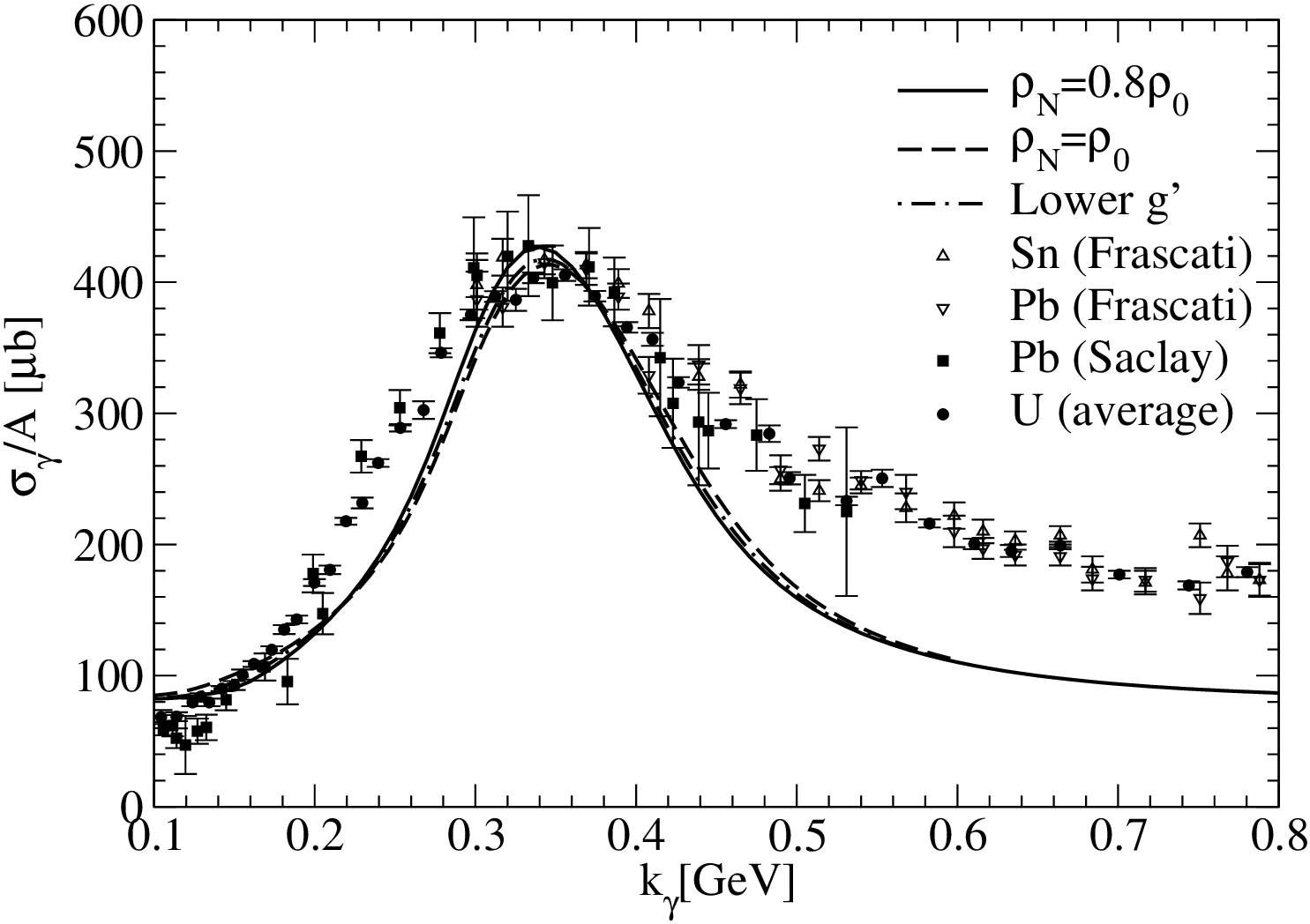}
\caption{Photoabsorption cross sections on 
 nuclei~\cite{ahr84,ahr85,fro92,fro94,bian93,bian96} compared
 to calculations employing our $\Delta$-spectral function at finite
 density. Solid line:
  $\varrho_N$=0.8$\varrho_0$, $g_{11}'$=0.8, $g_{12}'$=$g_{22}'$=0.33,
  dashed line: $\varrho_N$=$\varrho_0$, $g_{11}'$=0.8,
  $g_{12}'$=$g_{22}'$=0.33, dash-dotted line: $\varrho_N$=$0.8\varrho_0$,
  $g_{11}'$=0.6, $g_{12}'$=$g_{22}'$=0.2. \label{fig_photo}}
\end{minipage}\hspace*{3mm}
\begin{minipage}[b][8.5cm][t]{0.45\textwidth}
\includegraphics[width=\textwidth]{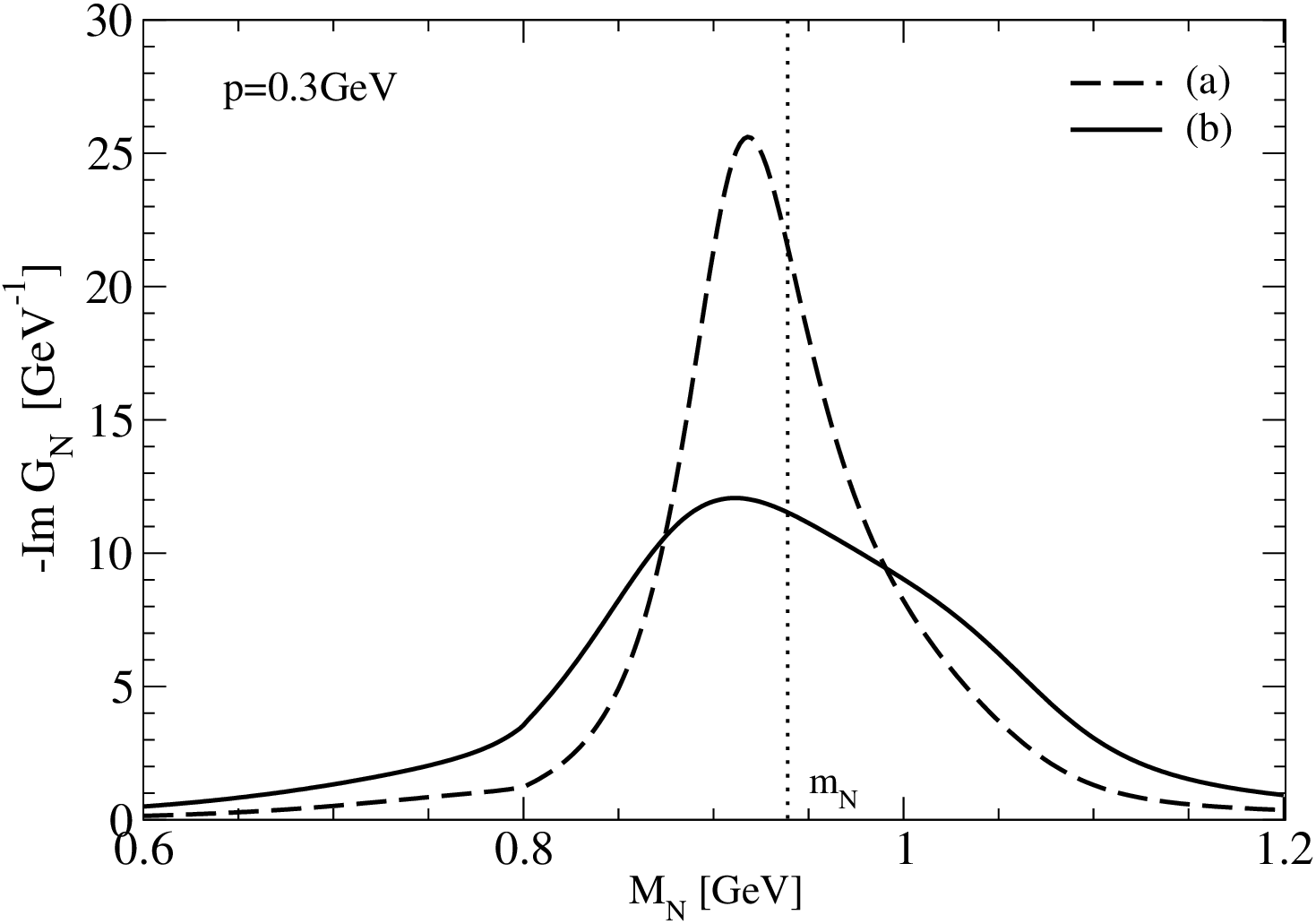}
\caption{Nucleon spectral function in hot hadronic matter under RHIC 
 conditions; solid line: ``chemical freezeout" with $T$=180MeV,
 $\varrho_N$=0.68$\varrho_0$ ($\mu_N$=333MeV), $\mu_{\pi}$=0; dashed
  line: ``thermal freezeout" with $T$=100MeV, $\varrho_N$=0.12$\varrho_0$
  ($\mu_N$=531MeV) and $\mu_{\pi}$=96MeV; the dotted line indicates the
  location of the free nucleon mass.\label{fig_nucl}}
\end{minipage}
\end{figure}
The photoabsorption cross section can be written in terms of the
photon-polarization tensor, $\Pi_\gamma$, as 
\begin{equation}
\frac{\sigma_{\gamma A}^{\mathrm{abs}}}{A}=\frac{4 \pi \alpha}{k}
\frac{1}{\varrho_N} \frac{1}{2} \im \Pi_{\gamma}(k_0=k)\ , \qquad 
\Pi_{\gamma}=\frac{1}{2} g_{\mu \nu} \Pi^{\mu \nu} \ ,
\label{sig_gam}
\end{equation}
where $k_0=k$ denotes the photon energy (momentum). $\Pi_\gamma$ is 
evaluated via the $\gamma$-induced $\Delta$-hole excitation which
we obtain from the standard magnetic coupling~\cite{ew88}
\begin{equation}
\Lag_{\gamma N \Delta}=-\frac{f_{\gamma N \Delta}}{4 \pi m_{\pi}}
\psi_N^{\dagger} (\vec{S}^{\dagger} \times \nabla) \vec{A} T_3^{\dagger}
\Psi_{\Delta} + \mathrm{h.c.}
\label{Lgnd}
\end{equation}
The cross section on the nucleon follows from the low-density limit of 
Eq.~(\ref{sig_gam}) using the vacuum $\Delta$ propagator (we also 
supplemented a non-resonant constant ``background" of 
80~$\mu$b~\cite{rubw98}); with a coupling constant 
$f_{\gamma N \Delta}$=0.653 and a (monopole-) form factor cutoff, 
$\Lambda_{\gamma N \Delta}$=400~MeV, the nucleon data in the 
$\Delta$-resonance region are resonably well reproduced~\cite{hr04b}.
The cross section on nuclei follows with no further adjustments 
by using the in-medium $\Delta$ propagator, assuming an average density
of $\rho_N$=0.8$\rho_0$ (the results are almost identical for 
$\rho_N$=$\rho_0$). Given our rather simple treatment, the agreement 
with the nuclear data is fair, cf. left panel of Fig.~\ref{fig_photo}. 
The sensitivity to changes in the Migdal parameters is very moderate.
The discrepancies at low energies may be due to neglecting
(i) interference between resonant and non-resonant terms, 
(ii) direct nucleon-hole excitations, or (iii) transverse contributions 
in the vertex corrections of the $\Delta$ decay with medium-modified 
$\rho$ mesons. Obviously, at higher energies, additional resonances need 
to be included.

%%%%%%%%%%%%%%%%%%%%%%%%%%%%%%%%%%%%%
\subsection{Hot Hadronic Matter}
\label{ssec_hot}
%%%%%%%%%%%%%%%%%%%%%%%%%%%%%%%%%%%%%%
We now turn to our main results, applying the model 
to hot and dense hadronic matter expected to be formed in 
heavy-ion collisions at RHIC (meson-dominated matter)  
and the future GSI facility (baryon-dominated matter).

The nucleon-spectral function is displayed in Fig.~\ref{fig_nucl} for RHIC
conditions. Resonant $\pi N$$\to$$B$ scattering is found to induce an
appreciable broadening (exceeding 200~MeV at the expected chemical
freezeout), which, around the free nucleon mass, is to $\sim$80\% due to
the $\Delta$ excitation (at $T$=180~MeV, the shoulder at somewhat higher
nucleon energies is mainly caused by the $N$(1520) and $\Delta$(1600)
resonances).  One also observes a slight attractive mass shift, induced by
the higher resonance states which are located (well) above typical $N \pi$
energies.

\begin{figure}[t]
\begin{minipage}{0.45\textwidth}
\includegraphics[width=0.95\textwidth]{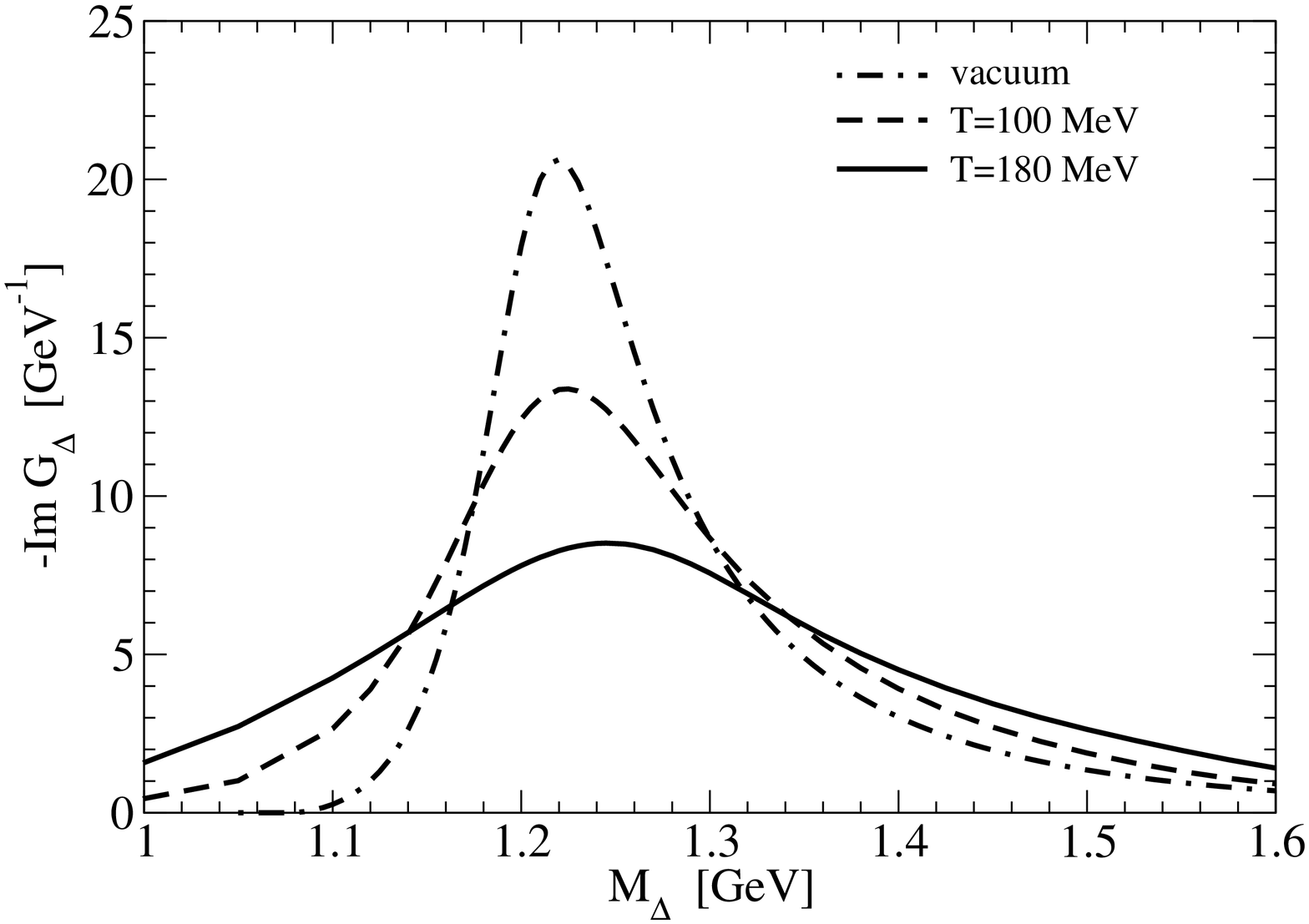}
\end{minipage}
\begin{minipage}{0.45\textwidth}
\includegraphics[width=0.95\textwidth]{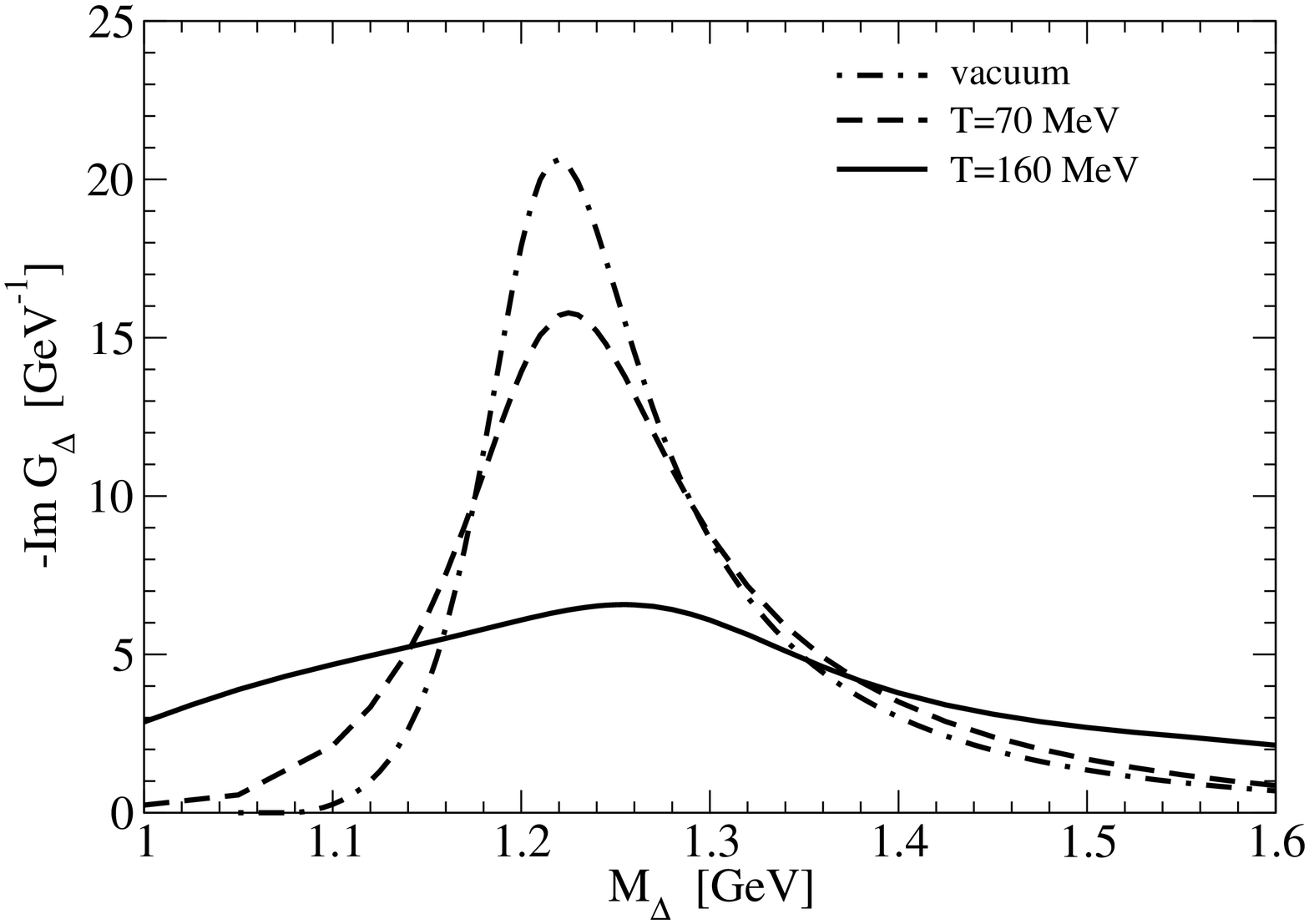}
\end{minipage}
\caption{Left: The in-medium spectral function of the $\Delta$(1232) for
  typical conditions at RHIC, compared to the vacuum (dash-dotted line).
  Dashed line: $T$=100~MeV, $\varrho_N$$=$0.12$\varrho_0$
  ($\mu_N$=531~MeV), $\mu_{\pi}$$=$96~MeV; solid line: $T$$=$180~MeV,
  $\varrho_N$$=$0.68$\varrho_0$ ($\mu_N=333 \; \text{MeV}$),
  $\mu_{\pi}$$=$0. Right: The same for typical conditions, expected at the
  future GSI facility.  Dashed line: $T$$=$70~MeV,
  $\varrho_N$$=$0.19$\varrho_0$ ($\mu_N$$=$727~MeV), $\mu_{\pi}$$=$105~MeV;
  solid line: $T$$=$160~MeV, $\varrho_N$$=$1.80$\varrho_0$
  ($\mu_N$$=$593~MeV), $\mu_{\pi}$$=$0.
\label{fig_del}}
\end{figure}

For the $\Delta(1232)$ spectral function at RHIC (left panel in 
Fig.~\ref{fig_del}), about half
of its width is due to baryon resonance excitations (which are 
additionally enhanced somewhat due to the inclusion of the in-medium
pion propagator in the loop diagram). The other half of the
in-medium broadening is mostly generated by the
Bose enhancement factor on the pion in the $\pi N$ decay.
In the real part of the $\Delta$ self-energy, there is a large
cancellation between the attraction generated by the baryon resonances
and the repulsion induced by the medium effects on the
$\pi N$ loop, mostly driven by finite baryon densities 
(with significant contributions from the vertex corrections).       

For conditions resembling thermal freezeout (dashed line in the left panel
of Fig.~\ref{fig_del}), the peak position is located at about
$M$$\simeq$1.226~GeV, and the line width has increased to
$\Gamma$$\simeq$177~MeV, to be compared to the vacuum values of
$M$$\simeq$1.219~GeV and $\Gamma$$\simeq$110~MeV, respectively.  The
in-medium changes in these quantities are in qualitative agreement with
preliminary data from the STAR collaboration~\cite{fach04a,zhang04}. For a
more conclusive comparison between our model and experimental data a
reliable description of the freezeout dynamics is necessary.

Towards the phase boundary, the width of the $\Delta$(1232) 
further increases substantially;  additional contributions from coupling
to in-medium $\rho$-mesons (as well as pertinent vertex corrections),
not included at present, are likely to accelerate the broadening
(without medium effects on the $\rho$, the $N \rho$ channel
is kinematically strongly suppressed).   

Finally, in the right panel of Fig.~\ref{fig_del}, we display
$\Delta$-spectral functions in a net-baryon rich environment, 
approximately corresponding to heavy-ion collisions at the future GSI
facility. Whereas around thermal freezeout the line shape is affected
rather little, close to the phase boundary the resonance structure
has essentially melted.
%%%%%%%%%%%%%%%%%%%%%%%%%%%%%%%%%%%%%%%
\section{Conclusions and Outlook}
\label{sec_concl}
%%%%%%%%%%%%%%%%%%%%%%%%%%%%%%%%%%%%%%%

Based on an effective hadronic model we have studied the properties of 
the nucleon and the $\Delta$(1232) in hot and dense matter. 
Medium modifications in the pion cloud of the $\Delta$ were accounted
for through in-medium pion and nucleon propagators (including vertex 
corrections), and pertinent thermal occupation factors. Direct 
interactions of $N$ and $\Delta$ with thermal pions have been 
approximated by baryon resonances, constrained
via empirical decay branchings. Within a simplified treatment we have 
checked that the nuclear effects on the pion lead to a $\Delta$ spectral
function in cold matter which provides fair agreement with 
nuclear photoabsorption.

Our main result is that, under RHIC conditions, the $\Delta$ ($N$) 
spectral function exhibits significant broadening and a slight upward 
(downward) peak shift. In the vicinity of thermal freezeout our findings 
are qualitatively in line with preliminary STAR data for $\pi N$
invariant-mass spectra, whereas at higher temperatures and densities
the broadening becomes more extreme, reminiscent to what has
been found for light vector mesons in the same framework~\cite{rw99}.     

Our work should also be considered as a step towards a comprehensive 
description of hadronic matter under extreme conditions, such
as its equation of state~\cite{vosk04}. 
Future developments will have to incorporate the coupling to vector 
mesons in a chiral framework, a more complete treatment of the pion and 
nucleon degrees of freedom (including nonresonant interactions), as 
well as medium effects on excited resonances.   
On the phenomenological side, to address $\pi N$ invariant-mass 
spectra in heavy-ion collisions, in-medium $\Delta$ spectral functions
need to be implemented into a dynamical description of the 
thermal freezeout. Furthermore, the large radiative decay branching 
of the $\Delta$ implies that its medium modifications
could play a role in electromagnetic emission spectra~\cite{Rapp04}, 
in particular the soft-photon enhancement recently observed in central 
Pb-Pb collisions at the SPS~\cite{agg03}.

\section*{Acknowledgment}

We thank J. Knoll, C.M. Ko, C. Korpa and F. Riek for discussions. 
H. van Hees acknowledges support from the Alexander-von-Humboldt 
Foundation as a Feodor-Lynen Fellow.

%\begin{flushleft}
%\bibliography{qftbib}

\begin{thebibliography}{41}
\expandafter\ifx\csname natexlab\endcsname\relax\def\natexlab#1{#1}\fi
\expandafter\ifx\csname bibnamefont\endcsname\relax
  \def\bibnamefont#1{#1}\fi
\expandafter\ifx\csname bibfnamefont\endcsname\relax
  \def\bibfnamefont#1{#1}\fi
\expandafter\ifx\csname citenamefont\endcsname\relax
  \def\citenamefont#1{#1}\fi
\expandafter\ifx\csname url\endcsname\relax
  \def\url#1{\texttt{#1}}\fi
\expandafter\ifx\csname urlprefix\endcsname\relax\def\urlprefix{URL }\fi
\providecommand{\bibinfo}[2]{#2}
\providecommand{\eprint}[2][]{\url{#2}}

\bibitem[{\citenamefont{Karsch}(2002)}]{kar01}
\bibinfo{author}{\bibfnamefont{F.}~\bibnamefont{Karsch}},
  \bibinfo{journal}{Lect. Notes Phys.} \textbf{\bibinfo{volume}{583}},
  \bibinfo{pages}{209} (\bibinfo{year}{2002}).

\bibitem[{\citenamefont{Rapp and Wambach}(2000)}]{rw99}
\bibinfo{author}{\bibfnamefont{R.}~\bibnamefont{Rapp}} \bibnamefont{and}
  \bibinfo{author}{\bibfnamefont{J.}~\bibnamefont{Wambach}},
  \bibinfo{journal}{Adv. Nucl. Phys.} \textbf{\bibinfo{volume}{25}},
  \bibinfo{pages}{1} (\bibinfo{year}{2000}).

\bibitem[{\citenamefont{Agakishiev et~al.}(1998)}]{ceres98}
\bibinfo{author}{\bibfnamefont{G.}~\bibnamefont{Agakishiev}}
  \bibnamefont{et~al.} (\bibinfo{collaboration}{CERES/NA45}),
  \bibinfo{journal}{Phys. Lett.} \textbf{\bibinfo{volume}{B422}},
  \bibinfo{pages}{405} (\bibinfo{year}{1998}).

\bibitem[{\citenamefont{Adamova et~al.}(2003)}]{ceres03}
\bibinfo{author}{\bibfnamefont{D.}~\bibnamefont{Adamova}} \bibnamefont{et~al.}
  (\bibinfo{collaboration}{CERES/NA45}), \bibinfo{journal}{Phys. Rev. Lett.}
  \textbf{\bibinfo{volume}{91}}, \bibinfo{pages}{042301}
  (\bibinfo{year}{2003}).

\bibitem[{\citenamefont{Hjort et~al.}(1997)}]{hjo97}
\bibinfo{author}{\bibfnamefont{E.~L.} \bibnamefont{Hjort}}
  \bibnamefont{et~al.}, \bibinfo{journal}{Phys. Rev. Lett.}
  \textbf{\bibinfo{volume}{79}}, \bibinfo{pages}{4345} (\bibinfo{year}{1997}).

\bibitem[{\citenamefont{Pelte et~al.}(1997)}]{pelte97}
\bibinfo{author}{\bibfnamefont{D.}~\bibnamefont{Pelte}} \bibnamefont{et~al.}
  (\bibinfo{collaboration}{FOPI}), \bibinfo{journal}{Z. Phys. A}
  \textbf{\bibinfo{volume}{359}}, \bibinfo{pages}{55} (\bibinfo{year}{1997}).

\bibitem[{\citenamefont{Zhang}(2004)}]{zhang04}
\bibinfo{author}{\bibfnamefont{H.-B.} \bibnamefont{Zhang}}
  (\bibinfo{collaboration}{STAR}) (\bibinfo{year}{2004}), Preprint:
  nucl-ex/0403010. 

\bibitem[{\citenamefont{Fachini}(2004)}]{fach04a}
\bibinfo{author}{\bibfnamefont{P.}~\bibnamefont{Fachini}}, \bibinfo{journal}{J.
  Phys. G} \textbf{\bibinfo{volume}{30}}, \bibinfo{pages}{S735}
  (\bibinfo{year}{2004}).

\bibitem[{\citenamefont{Horikawa et~al.}(1980)\citenamefont{Horikawa, Thies,
  and Lenz}}]{hori80}
\bibinfo{author}{\bibfnamefont{Y.}~\bibnamefont{Horikawa}},
  \bibinfo{author}{\bibfnamefont{M.}~\bibnamefont{Thies}}, \bibnamefont{and}
  \bibinfo{author}{\bibfnamefont{F.}~\bibnamefont{Lenz}},
  \bibinfo{journal}{Nucl. Phys.} \textbf{\bibinfo{volume}{A345}},
  \bibinfo{pages}{386} (\bibinfo{year}{1980}).

\bibitem[{\citenamefont{Oset and Salcedo}(1987)}]{os87}
\bibinfo{author}{\bibfnamefont{E.}~\bibnamefont{Oset}} \bibnamefont{and}
  \bibinfo{author}{\bibfnamefont{L.~L.} \bibnamefont{Salcedo}},
  \bibinfo{journal}{Nucl. Phys.} \textbf{\bibinfo{volume}{A468}},
  \bibinfo{pages}{631} (\bibinfo{year}{1987}).

\bibitem[{\citenamefont{Ericson and Weise}(1988)}]{ew88}
\bibinfo{author}{\bibfnamefont{T.}~\bibnamefont{Ericson}} \bibnamefont{and}
  \bibinfo{author}{\bibfnamefont{W.}~\bibnamefont{Weise}},
  \emph{\bibinfo{title}{{Pions} and {Nuclei}}} (\bibinfo{publisher}{Clarendon
  Press, Oxford}, \bibinfo{year}{1988}).

\bibitem[{\citenamefont{Migdal et~al.}(1990)\citenamefont{Migdal, Saperstein,
  Troitsky, and Voskresensky}}]{mig90}
\bibinfo{author}{\bibfnamefont{A.~B.} \bibnamefont{Migdal}},
  \bibinfo{author}{\bibfnamefont{E.~E.} \bibnamefont{Saperstein}},
  \bibinfo{author}{\bibfnamefont{M.~A.} \bibnamefont{Troitsky}},
  \bibnamefont{and} \bibinfo{author}{\bibfnamefont{D.~N.}
  \bibnamefont{Voskresensky}}, \bibinfo{journal}{Phys. Rept.}
  \textbf{\bibinfo{volume}{192}}, \bibinfo{pages}{179} (\bibinfo{year}{1990}).

\bibitem[{\citenamefont{Xia et~al.}(1994)\citenamefont{Xia, Siemens, and
  Soyeur}}]{xia94}
\bibinfo{author}{\bibfnamefont{L.~H.} \bibnamefont{Xia}},
  \bibinfo{author}{\bibfnamefont{P.~J.} \bibnamefont{Siemens}},
  \bibnamefont{and} \bibinfo{author}{\bibfnamefont{M.}~\bibnamefont{Soyeur}},
  \bibinfo{journal}{Nucl. Phys.} \textbf{\bibinfo{volume}{A578}},
  \bibinfo{pages}{493} (\bibinfo{year}{1994}).

\bibitem[{\citenamefont{Korpa and Lutz}(2004)}]{lutz03}
\bibinfo{author}{\bibfnamefont{C.~L.} \bibnamefont{Korpa}} \bibnamefont{and}
  \bibinfo{author}{\bibfnamefont{M.~F.~M.} \bibnamefont{Lutz}},
  \bibinfo{journal}{Nucl. Phys.} \textbf{\bibinfo{volume}{A742}},
  \bibinfo{pages}{305} (\bibinfo{year}{2004}).

\bibitem[{\citenamefont{Ko et~al.}(1989)\citenamefont{Ko, Xia, and
  Siemens}}]{ko89}
\bibinfo{author}{\bibfnamefont{C.~M.} \bibnamefont{Ko}},
  \bibinfo{author}{\bibfnamefont{L.~H.} \bibnamefont{Xia}}, \bibnamefont{and}
  \bibinfo{author}{\bibfnamefont{P.~J.} \bibnamefont{Siemens}},
  \bibinfo{journal}{Phys. Lett.} \textbf{\bibinfo{volume}{B231}},
  \bibinfo{pages}{16} (\bibinfo{year}{1989}).

\bibitem[{\citenamefont{Korpa and Malfliet}(1995)}]{korpa95}
\bibinfo{author}{\bibfnamefont{C.~L.} \bibnamefont{Korpa}} \bibnamefont{and}
  \bibinfo{author}{\bibfnamefont{R.}~\bibnamefont{Malfliet}},
  \bibinfo{journal}{Phys. Rev. C} \textbf{\bibinfo{volume}{52}},
  \bibinfo{pages}{2756} (\bibinfo{year}{1995}).

\bibitem[{\citenamefont{Leutwyler and Smilga}(1990)}]{LS90}
\bibinfo{author}{\bibfnamefont{H.}~\bibnamefont{Leutwyler}} \bibnamefont{and}
  \bibinfo{author}{\bibfnamefont{A.~V.} \bibnamefont{Smilga}},
  \bibinfo{journal}{Nucl. Phys.} \textbf{\bibinfo{volume}{B342}},
  \bibinfo{pages}{302} (\bibinfo{year}{1990}).

\bibitem[{\citenamefont{Dominguez et~al.}(1994)\citenamefont{Dominguez, Loewe,
  and Rojas}}]{DLR94}
\bibinfo{author}{\bibfnamefont{C.~A.} \bibnamefont{Dominguez}},
  \bibinfo{author}{\bibfnamefont{M.}~\bibnamefont{Loewe}}, \bibnamefont{and}
  \bibinfo{author}{\bibfnamefont{J.~C.} \bibnamefont{Rojas}},
  \bibinfo{journal}{Phys. Lett.} \textbf{\bibinfo{volume}{B320}},
  \bibinfo{pages}{377} (\bibinfo{year}{1994}).

\bibitem[{\citenamefont{Eletsky and Ioffe}(1997)}]{EI97}
\bibinfo{author}{\bibfnamefont{V.~L.} \bibnamefont{Eletsky}} \bibnamefont{and}
  \bibinfo{author}{\bibfnamefont{B.~L.} \bibnamefont{Ioffe}},
  \bibinfo{journal}{Phys. Lett.} \textbf{\bibinfo{volume}{B401}},
  \bibinfo{pages}{327} (\bibinfo{year}{1997}).

\bibitem[{\citenamefont{Kacir and Zahed}(1996)}]{KZ96}
\bibinfo{author}{\bibfnamefont{M.}~\bibnamefont{Kacir}} \bibnamefont{and}
  \bibinfo{author}{\bibfnamefont{I.}~\bibnamefont{Zahed}},
  \bibinfo{journal}{Phys. Rev. D} \textbf{\bibinfo{volume}{54}},
  \bibinfo{pages}{5536} (\bibinfo{year}{1996}).

\bibitem[{\citenamefont{Cubero}(1990)}]{cub90}
\bibinfo{author}{\bibfnamefont{M.}~\bibnamefont{Cubero}}, Ph.D. thesis,
  \bibinfo{school}{TH Darmstadt} (\bibinfo{year}{1990}).

\bibitem[{\citenamefont{Rapp et~al.}(1998)\citenamefont{Rapp, Urban, Buballa,
  and Wambach}}]{rubw98}
\bibinfo{author}{\bibfnamefont{R.}~\bibnamefont{Rapp}},
  \bibinfo{author}{\bibfnamefont{M.}~\bibnamefont{Urban}},
  \bibinfo{author}{\bibfnamefont{M.}~\bibnamefont{Buballa}}, \bibnamefont{and}
  \bibinfo{author}{\bibfnamefont{J.}~\bibnamefont{Wambach}},
  \bibinfo{journal}{Phys. Lett.} \textbf{\bibinfo{volume}{B417}},
  \bibinfo{pages}{1} (\bibinfo{year}{1998}).

\bibitem[{\citenamefont{Urban et~al.}(2000)\citenamefont{Urban, Buballa, Rapp,
  and Wambach}}]{ubw99}
\bibinfo{author}{\bibfnamefont{M.}~\bibnamefont{Urban}},
  \bibinfo{author}{\bibfnamefont{M.}~\bibnamefont{Buballa}},
  \bibinfo{author}{\bibfnamefont{R.}~\bibnamefont{Rapp}}, \bibnamefont{and}
  \bibinfo{author}{\bibfnamefont{J.}~\bibnamefont{Wambach}},
  \bibinfo{journal}{Nucl. Phys.} \textbf{\bibinfo{volume}{A673}},
  \bibinfo{pages}{357} (\bibinfo{year}{2000}).

\bibitem[{\citenamefont{Nacher et~al.}(2001)\citenamefont{Nacher, Oset,
  Vicente, and Roca}}]{novr01}
\bibinfo{author}{\bibfnamefont{J.~C.} \bibnamefont{Nacher}},
  \bibinfo{author}{\bibfnamefont{E.}~\bibnamefont{Oset}},
  \bibinfo{author}{\bibfnamefont{M.~J.} \bibnamefont{Vicente}},
  \bibnamefont{and} \bibinfo{author}{\bibfnamefont{L.}~\bibnamefont{Roca}},
  \bibinfo{journal}{Nucl. Phys.} \textbf{\bibinfo{volume}{A695}},
  \bibinfo{pages}{295} (\bibinfo{year}{2001}).

\bibitem[{\citenamefont{Hagiwara et~al.}(2002)}]{pdb02}
\bibinfo{author}{\bibfnamefont{K.}~\bibnamefont{Hagiwara}}
  \bibnamefont{et~al.}, \bibinfo{journal}{Phys. Rev. D}
  \textbf{\bibinfo{volume}{66}}, \bibinfo{pages}{01001} (\bibinfo{year}{2002}).

\bibitem[{\citenamefont{Moniz}(1981)}]{mon80}
\bibinfo{author}{\bibfnamefont{E.~J.} \bibnamefont{Moniz}},
  \bibinfo{journal}{Nucl. Phys.} \textbf{\bibinfo{volume}{A354}},
  \bibinfo{pages}{535c} (\bibinfo{year}{1981}).

\bibitem[{\citenamefont{Weinhold et~al.}(1998)\citenamefont{Weinhold, Friman,
  and N{\"o}renberg}}]{wfn98}
\bibinfo{author}{\bibfnamefont{W.}~\bibnamefont{Weinhold}},
  \bibinfo{author}{\bibfnamefont{B.}~\bibnamefont{Friman}}, \bibnamefont{and}
  \bibinfo{author}{\bibfnamefont{W.}~\bibnamefont{N{\"o}renberg}},
  \bibinfo{journal}{Phys. Lett.} \textbf{\bibinfo{volume}{B433}},
  \bibinfo{pages}{236} (\bibinfo{year}{1998}).

\bibitem[{\citenamefont{Weinhold}(1995)}]{weinh95}
\bibinfo{author}{\bibfnamefont{W.}~\bibnamefont{Weinhold}}, Diploma thesis,
  \bibinfo{school}{TH Darmstadt} (\bibinfo{year}{1995}).

\bibitem[{\citenamefont{van Hees and Knoll}(2002)}]{vHK2001-Ren-II}
\bibinfo{author}{\bibfnamefont{H.}~\bibnamefont{van Hees}} \bibnamefont{and}
  \bibinfo{author}{\bibfnamefont{J.}~\bibnamefont{Knoll}},
  \bibinfo{journal}{Phys. Rev. D} \textbf{\bibinfo{volume}{65}},
  \bibinfo{pages}{105005} (\bibinfo{year}{2002}).

\bibitem[{\citenamefont{Rapp and Wambach}(1995)}]{rw95}
\bibinfo{author}{\bibfnamefont{R.}~\bibnamefont{Rapp}} \bibnamefont{and}
  \bibinfo{author}{\bibfnamefont{J.}~\bibnamefont{Wambach}},
  \bibinfo{journal}{Phys. Lett.} \textbf{\bibinfo{volume}{B351}},
  \bibinfo{pages}{50} (\bibinfo{year}{1995}).

\bibitem[{\citenamefont{Migdal}(1978)}]{mig78}
\bibinfo{author}{\bibfnamefont{A.~B.} \bibnamefont{Migdal}},
  \bibinfo{journal}{Rev. Mod. Phys.} \textbf{\bibinfo{volume}{50}},
  \bibinfo{pages}{107} (\bibinfo{year}{1978}).

\bibitem[{\citenamefont{Chanfray and Schuck}(1993)}]{cs93}
\bibinfo{author}{\bibfnamefont{G.}~\bibnamefont{Chanfray}} \bibnamefont{and}
  \bibinfo{author}{\bibfnamefont{P.}~\bibnamefont{Schuck}},
  \bibinfo{journal}{Nucl. Phys.} \textbf{\bibinfo{volume}{A555}},
  \bibinfo{pages}{329} (\bibinfo{year}{1993}).

\bibitem[{\citenamefont{Herrmann et~al.}(1993)\citenamefont{Herrmann, Friman,
  and N{\"o}renberg}}]{hfn93}
\bibinfo{author}{\bibfnamefont{M.}~\bibnamefont{Herrmann}},
  \bibinfo{author}{\bibfnamefont{B.~L.} \bibnamefont{Friman}},
  \bibnamefont{and}
  \bibinfo{author}{\bibfnamefont{W.}~\bibnamefont{N{\"o}renberg}},
  \bibinfo{journal}{Nucl. Phys.} \textbf{\bibinfo{volume}{A560}},
  \bibinfo{pages}{411} (\bibinfo{year}{1993}).

\bibitem[{\citenamefont{Braun-Munzinger
  et~al.}(2003)\citenamefont{Braun-Munzinger, Redlich, and Stachel}}]{pbm03}
\bibinfo{author}{\bibfnamefont{P.}~\bibnamefont{Braun-Munzinger}},
  \bibinfo{author}{\bibfnamefont{K.}~\bibnamefont{Redlich}}, \bibnamefont{and}
  \bibinfo{author}{\bibfnamefont{J.}~\bibnamefont{Stachel}}
  (\bibinfo{year}{2003}), Preprint: nucl-th/0304013.

\bibitem[{\citenamefont{Rapp}(2002)}]{rap02}
\bibinfo{author}{\bibfnamefont{R.}~\bibnamefont{Rapp}}, \bibinfo{journal}{Phys.
  Rev. C} \textbf{\bibinfo{volume}{66}}, \bibinfo{pages}{017901}
  (\bibinfo{year}{2002}).

\bibitem[{\citenamefont{Korpa and Dieperink}(2004)}]{korpa04}
\bibinfo{author}{\bibfnamefont{C.~L.} \bibnamefont{Korpa}} \bibnamefont{and}
  \bibinfo{author}{\bibfnamefont{A.~E.~L.} \bibnamefont{Dieperink}},
  \bibinfo{journal}{Phys. Rev.} \textbf{\bibinfo{volume}{C70}},
  \bibinfo{pages}{015207} (\bibinfo{year}{2004}).

\bibitem[{\citenamefont{Ahrens et~al.}(1984)}]{ahr84}
\bibinfo{author}{\bibfnamefont{J.}~\bibnamefont{Ahrens}} \bibnamefont{et~al.},
  \bibinfo{journal}{Phys. Lett.} \textbf{\bibinfo{volume}{B146}},
  \bibinfo{pages}{303} (\bibinfo{year}{1984}).

\bibitem[{\citenamefont{Ahrens}(1985)}]{ahr85}
\bibinfo{author}{\bibfnamefont{J.}~\bibnamefont{Ahrens}},
  \bibinfo{journal}{Nucl. Phys.} \textbf{\bibinfo{volume}{A446}},
  \bibinfo{pages}{229c} (\bibinfo{year}{1985}).

\bibitem[{\citenamefont{Frommhold et~al.}(1992)}]{fro92}
\bibinfo{author}{\bibfnamefont{T.}~\bibnamefont{Frommhold}}
  \bibnamefont{et~al.}, \bibinfo{journal}{Phys. Lett.}
  \textbf{\bibinfo{volume}{B295}}, \bibinfo{pages}{28} (\bibinfo{year}{1992}).

\bibitem[{\citenamefont{Frommhold et~al.}(1994)}]{fro94}
\bibinfo{author}{\bibfnamefont{T.}~\bibnamefont{Frommhold}}
  \bibnamefont{et~al.}, \bibinfo{journal}{Z. Phys. A}
  \textbf{\bibinfo{volume}{350}}, \bibinfo{pages}{249} (\bibinfo{year}{1994}).

\bibitem[{\citenamefont{Bianchi et~al.}(1993)}]{bian93}
\bibinfo{author}{\bibfnamefont{N.}~\bibnamefont{Bianchi}} \bibnamefont{et~al.},
  \bibinfo{journal}{Phys. Lett.} \textbf{\bibinfo{volume}{B299}},
  \bibinfo{pages}{219} (\bibinfo{year}{1993}).

\bibitem[{\citenamefont{Bianchi et~al.}(1996)}]{bian96}
\bibinfo{author}{\bibfnamefont{N.}~\bibnamefont{Bianchi}} \bibnamefont{et~al.},
  \bibinfo{journal}{Phys. Rev. C} \textbf{\bibinfo{volume}{54}},
  \bibinfo{pages}{1688} (\bibinfo{year}{1996}).

\bibitem[{\citenamefont{van Hees and Rapp}(2004)}]{hr04b}
\bibinfo{author}{\bibfnamefont{H.}~\bibnamefont{van Hees}} \bibnamefont{and}
  \bibinfo{author}{\bibfnamefont{R.}~\bibnamefont{Rapp}}
  (\bibinfo{year}{2004}), Preprint: nucl-th/0409026.

\bibitem[{\citenamefont{Voskresensky}(2004)}]{vosk04}
\bibinfo{author}{\bibfnamefont{D.~N.} \bibnamefont{Voskresensky}}
  (\bibinfo{year}{2004}), Preprint: hep-ph/0402020.

\bibitem[{\citenamefont{Rapp}(2004)}]{Rapp04}
\bibinfo{author}{\bibfnamefont{R.}~\bibnamefont{Rapp}}, \bibinfo{journal}{Mod.
  Phys. Lett.} \textbf{\bibinfo{volume}{A19}}, \bibinfo{pages}{1717}
  (\bibinfo{year}{2004}).

\bibitem[{\citenamefont{Aggarwal et~al.}(2004)}]{agg03}
\bibinfo{author}{\bibfnamefont{M.~M.} \bibnamefont{Aggarwal}}
  \bibnamefont{et~al.} (\bibinfo{collaboration}{WA98}), \bibinfo{journal}{Phys.
  Rev. Lett.} \textbf{\bibinfo{volume}{93}}, \bibinfo{pages}{022301}
  (\bibinfo{year}{2004}).

\end{thebibliography}
%\end{flushleft}

\end{document}